\documentclass[12pt]{article}
\usepackage{amssymb,amsmath,amsthm,latexsym}

\newtheorem{thm}{Theorem}[section]
\newtheorem{prop}[thm]{Proposition}
\newtheorem{lem}[thm]{Lemma}
\newtheorem{cor}[thm]{Corollary}
\newtheorem{exam}{Example}
\newtheorem{defi}[thm]{Definition}
\newtheorem{rem}[thm]{Remark}
\newcommand{\pf}{{\bf Proof. \ }}

\begin{document}

\title{Quantum Codes from Linear Codes over Finite Chain Rings}
\author{
 Xiusheng Liu\\
 School of Mathematics and Physics, \\
 Hubei Polytechnic University  \\
 Huangshi, Hubei 435003, China, \\
{Email: \tt lxs6682@163.com} \\
Hualu Liu\\
 School of Mathematics and Physics, \\
 Hubei Polytechnic University  \\
 Huangshi, Hubei 435003, China, \\
{Email: \tt hwlulu@aliyun.com} \\}
\maketitle


\begin{abstract} In this paper, we provide two methods of constructing  quantum codes from linear codes over finite chain rings.  The first one is derived from the Calderbank-Shor-Steane (CSS) construction applied to self-dual codes over finite chain rings. The second construction is derived from the CSS construction applied to Gray images of the linear codes over finite chain ring $\mathbb{F}_{p^{2m}}+u\mathbb{F}_{p^{2m}}$.  The good parameters of quantum codes from cyclic codes over finite chain rings are obtained.

\end{abstract}


\bf Keywords\rm ~ Quantum  codes $\cdot$  Cyclic codes $\cdot$ Dual-containing codes,

\section{Introduction}
Quantum codes were introduced to protect quantum information from decoherence and quantum noise. A main obstacle to complete quantum communication is decoherence of quantum bits caused by inevitable interaction with environments. Quantum codes provide an efficient way to overcome decoherence. After the works of Shor [1] and Steane [2], the theory of quantum codes has been progressed rapidly.
Calderbank et al. [3] provided a systematic mathematical scheme to construct quantum codes from classical Hermitian dual-containing codes over finite field $\mathbb{F}_{4}$.

Some authors constructed quantum codes by using linear codes over finite rings. For example, in [4], Qian et al. gave a new method to obtain self-orthogonal codes over $\mathbb{F}_{2}$. They gave a construction for quantum codes starting from cyclic codes over finite ring, $\mathbb{F}_{2} +u\mathbb{F}_{2}, u^{2} = 0$.   In [5],  Ashraf and Mohammad  constructed
quantum codes from cyclic codes over $\mathbb{F}_3 + v\mathbb{F}_3$. In this paper, we continue to study quantum codes which are derived from  finite  chain ring.

Hereafter, $p$ is a prime. The purpose of this paper is to consider liner codes over finite chain rings to obtain good quantum codes.
In Section 2, we review some concepts and properties about quantum codes over finite fields.
In section 3,   we first give the necessary background materials on finite chain rings. Then a construction for quantum codes from self-dual codes over finite chain rings is given.  In the final section, for special finite chain ring $R=\mathbb{F}_{p^{2m}}+u\mathbb{F}_{p^{2m}}$, we define a new Gray map $\Phi$ from $R^n$ to $\mathbb{F}_{p^{2m}}^{2n}$, Gray weights of elements of $R^n$,  Gray distance $d_G(C)$ and  Hermitian dual $C^{\perp_{H}}$ with respect to Gray weight and the Hermitian inner product in the linear code $C$  over $R$. We prove that $\Phi(C^{\perp_{H}})=\Phi(C)^{\perp_{H}}$, and give a method to derive Hermitian dual-containing codes over  $\mathbb{F}_{p^{2m}}$ as Gray images of linear codes over $\mathbb{F}_{p^{2m}}+u\mathbb{F}_{p^{2m}}$. The parameters of quantum codes are obtained from cyclic codes over $R$.

\section{Review of Symmetric Quantum Codes}

In this section, we recall some basic concepts and results of symmetric quantum codes, necessary for the development of this work. For more details, we refer to [6,7].

Let $\mathbb{F}_{q}$ be the finite field with $q=p^{2m}$, where $p$ is a prime number and $m\geq1$ is an integer. Let $V_n$ be the Hilbert space $V_n=\mathbb{C}^{q^n}=\mathbb{C}^{q}\otimes\cdots\otimes\mathbb{C}^{q}$. Let $ \mid x\rangle$ be the vectors of an orthonormal basis of $\mathbb{C}^{q^n}$, where the labels $x$ are elements of $\mathbb{F}_{q}$. Then $V_n$ has the following orthonormal basis $\{|c\rangle=|c_1c_2\cdots c_n\rangle=|c_1\rangle \otimes|c_2\rangle \otimes\cdots\otimes| c_n\rangle:~c=(c_1,c_2,\ldots,c_n) \in \mathbb{F}_{q}^{n}\}$.

Consider $a,b\in \mathbb{F}_{q}$, the unitary linear operators $X(a)$ and $Z(b)$ in $\mathbb{C}^{q}$ are defined by $X(a)|x\rangle=|x+a\rangle$ and $Z(b)|x\rangle=w^{tr(bx)}|x\rangle$, respectively, where $w=exp(2\pi i/p)$ is a primitive $p$-th root of unity and  tr is the trace map from  $\mathbb{F}_{q}$ to  $\mathbb{F}_{p}$.

Let $\mathbf{a}=(a_1,\ldots,a_n)\in \mathbb{F}_{q}^n$, we write $X(\mathbf{a})=X(a_1)\otimes\cdots\otimes X(a_n)$ and $Z(\mathbf{a})=Z(a_1)\otimes\cdots\otimes Z(a_n)$ for the tensor products of $n$ error operators. The set $E_n=\{X(\mathbf{a})Z(\mathbf{b}): \mathbf{a},\mathbf{b}\in \mathbb{F}_{q}^n\}$ is an error basis on the complex vector space $\mathbb{C}^{q^n}$ and  we set $G_n=\{w^cX(\mathbf{a})Z(\mathbf{b}): \mathbf{a},\mathbf{b}\in \mathbb{F}_{q}^n,c\in\mathbb{F}_{p} \}$ is the error group associated with $E_n$.

\begin{defi}
 A $q$-ary quantum code of length $n$ is a subspace $Q$ of $V_n$ with dimension $K>1$.  A quantum code $Q$ of dimension $K>2$ is called symmetric quantum code (SQC) with parameters $((n,K,d))_q$ or $[[n,k,d]]_q$,where $k=log_qK$ if $Q$ detect $d-1$ quantum digits of errors for $d\geq1$. Namely, if for every orthogonal pair $|u\rangle,|v\rangle$ in $Q$ with $<u|v>=0$ and every $e\in G_n$ with $W_Q(e)\leq d-1$, $|u\rangle$ and $e|v\rangle$ are orthogonal, i.e.,$<u|e|v>=0$. Such a quantum code is called pure if $<u|e|v>=0$ for any  $|u\rangle$ and $|v\rangle$  in $Q$ and any $e\in G_n$ with $1\leq W_Q(e)\leq d-1$. A quantum code $Q$ with $K=1$ is always pure.
\end{defi}
Let us recall the SQC construction:
\begin{thm}
$[7,\mathrm{Lemma}~20]$ Let $C_i  $ be a classical linear code with parameters $[n,k_i,d_i]_{q}$ and $C_{i}^{\perp}\subseteq C_{1+(i~mod~2)}$$(i=1,2)$. Then there exists an SQC  Q  with parameters $[[n,k_1+k_2-n,\geq d]]_{q}$ that is pure to $min \{d_1,d_2\}$, where $d=min\{wt(c):c\in(C_1\backslash C_2^{\perp})\cup(C_2\backslash C_1^{\perp})\}$.
\end{thm}
\begin{cor}
If $C$  be  a classical linear $[n,k,d]_{q}$ code containing its dual $C^\perp\subseteq C$, then there exists an SQC  Q  with parameters $[[n,2k-n,\geq d]]_{q}$ that is pure to $d$.
\end{cor}

To see that an SQC  is good in terms of its parameters, we have to introduce the quantum Singleton bound ( See [6]).
\begin{thm}
Let $C$ be an SQC with parameters $[[n,k,d]]_{q}$. Then $k\leq n-2d+2$.
\end{thm}

If  an  SQC  $Q$ with parameters $[[n,k,d]]_{q}$ attains the quantum Singleton bound $k= n-2d+2$, then it is called an SQC maximum-distance-separable (SQCMDS) code.
\begin{defi} An SQC  $Q$ with parameters $[[n,k,d]]_{q}$ is called a near quantum maximum
distance separable (SQCNMDS) code if it satisfies $2d \geq n - k$.
\end{defi}
\begin{cor}\label{corollary4.8}
Let $C$ be an $[n,k,d]_{q}$ classical  code containing its dual, $C^\perp\subseteq C$. Then
\begin{itemize}
\item[{(1)}] $k\geq\lceil\frac{n}{2}\rceil.$

\item[{(2)}] If $C$ is an MDS code, then there exists an  $[[n,2k-n,d]]_{q}$ SQCMDS code.
\end{itemize}
\end{cor}
\pf(1)  Since $C$ is a $k$-dimensional subspace of  $\mathbb{F}_{q}^n$,  $C^\perp$ is a $(n-k)$-dimensional subspace of  $\mathbb{F}_{q}^n$. It follows  that $n-k\leq k$ by $C^\perp\subseteq C$. Therefore,  $k\geq\lceil\frac{n}{2}\rceil.$

(2) If $C$ is an $[n,k,d]_{q}$ classical MDS codes  containing its dual,$C^\perp\subseteq C$, then Corollary 2.3  implies the existence of a quantum
$[[n,2k-n,\geq d]]_{q}$ code $Q$. Theorem 2.4 shows that the minimum distance of $Q$ is $\leq\frac{n-(2k-n)+2}{2}=n-k+1$, so $Q$ is an   $[[n,2k-n,d]]_{q}$ SQCMDS code.

\begin{cor}\label{corollary4.8}
Let $C$ be an $[n,k,d]_{q}$ classical code containing its dual, $C^\perp\subseteq C$, and $2d\geq n-k $. Then
there exists an  $[[n,2k-n,\geq d]]_{q}$ SQCNMDS code.

\end{cor}
\pf If $C$ is an $[n,k,d]_{q}$ classical codes  containing its dual,$C^\perp\subseteq C$, then Corollary 2.3  implies the existence of a quantum
$[[n,2k-n,d_1]]_{q}$ code $Q$, where $d_1\geq d$. So $Q$ is an   $[[n,2k-n,\geq d]]_{q}$ SQCNMDS code.
\section{SQC  from to the Linear Codes over Finite Chain Rings}

Constructions of quantum codes are exhaustively investigated in the literature. As mentioned in Section 1, some authors have exhibited families of optimal codes. However, many of these techniques are based on the construction of classical codes over finite fields.

In this section, we use self-dual codes over finite chain rings to construct SQC.

We begin with some definitions and properties about finite chain rings (see[8,9]).
Let $R$ be a finite commutative ring with identity. A nonempty subset $C\subseteq R^{n}$ is called a {\it linear code} of length $n$ over $R$ if it is an $R$-submodule of $R^{n}$. Throughout this section,   all codes are assumed to be linear.

A commutative ring is called a {\it chain ring} if the lattice of all its ideals is a chain.
It is well known that if $R$ is a finite chain ring, then $R$ is a principal ideal ring and
has a unique maximal ideal $\langle \gamma\rangle=R\gamma=\{r\gamma\,|\,r\in R\}$. Its chain of ideals is

\begin{equation*}
R=\langle\gamma^{0}\rangle\supset\langle\gamma^{1}\rangle\supset\cdots\langle\gamma^{t-1}\rangle\supset\langle\gamma^{t}\rangle=\{0\}.
\end{equation*}
The integer $t$ is called  the {\it nilpotency index} of $\gamma$.
Note that the quotient $R/{\langle\gamma\rangle}$
is a finite field $\mathbb{F}_q$,
where $q$ is a power of a prime $p$.
There is a natural homomorphism from $R$ onto  $\mathbb{F}_q=R/\langle\gamma\rangle$, i.e.,
$$
^{-}:R\longrightarrow \mathbb{F}_q=R/\langle\gamma\rangle,~~~r\mapsto r+\langle\gamma\rangle=\overline{r},
~~~~\hbox{for any $r\in R$}.
$$

We need the following lemma (see [9]).
\begin{lem}
 Let $R$ be a finite chain ring with maximal ideal $\langle\gamma\rangle$. Let $V\subset R$ be a set of representatives for the equivalence classes of $R$ under congruence modulo $\gamma$. Then
\begin{itemize}
\item[{(1)}]for any $v\in R$ there exist unique $v_{0},\ldots,v_{t-1}\in V$ such that $v=\sum_{i=0}^{t-1}v_{i}\gamma^{i}.$

\item[{(2)}]$|V|=|R/\gamma|=|\mathbb{F}_q|$.
\end{itemize}
\end{lem}

The natural homomorphism from $R$ onto  $\mathbb{F}_q=R/\langle\gamma\rangle$ can be extended naturally to a projection from $R^{n}$ onto $\mathbb{F}_q^{n}$. For an element $c\in R^{n}$, let $\overline{c}$ be its image under this projection.
Given    $r\in R$ and   $c\in R^n$, we denote by $rc$ the usual multiplication of a vector by a scalar.
Let $C$ be a   code of length $n$ over $R$.
We define $\overline{C}=\big\{\overline{c}\,|\,c\in C\big\}$ and $(C:r)=\big\{e\in R^{n}\,|\,re\in C\big\}$,
where  $r$ is an element of $R$.
\begin{defi}
To any code $C$ over $R$, we associate the tower of codes
$$C=(C:\gamma^{0})\subseteq(C:\gamma)\subseteq\ldots \subseteq(C:\gamma^{t-1})$$
over $R$ and its projection to $\mathbb{F}_q$,
$$\overline{C}=\overline{(C:\gamma^{0})}\subseteq\overline{(C:\gamma)}\subseteq\ldots \subseteq\overline{(C:\gamma^{t-1})}.$$
\end{defi}

The following definitions and  results can be found in \cite{Norton}.
\begin{defi}
Let $C$ be a linear code over $R$. A generator matrix $G$ for $C$ is said to be in standard form if after a suitable permutation of the coordinates,
$$G=\begin{pmatrix}I_{k_{0}} & A_{0,1} & A_{0,2} & A_{0,3} & \cdots &  A_{0,t-1} & A_{0,t} \\
     0 & \gamma I_{k_{1}} &\gamma A_{1,2} &\gamma A_{1,3} & \cdots & \gamma A_{1,t-1} & \gamma A_{1,t} \\
    0 & 0 & \gamma^2 I_{k_{2}} & \gamma^2 A_{2,3} & \cdots & \gamma^2 A_{2,t-1} & \gamma^2 A_{2,t} \\
    \cdots& \cdots & \cdots & \cdots & \cdots & \cdots & \cdots \\
    0 & 0 & 0 & 0 & \cdots & \gamma^{t-1} I_{k_{t-1}} & \gamma^{t-1} A_{t-1,t} \\
  \end{pmatrix}
=\begin{pmatrix}
A_{0}\\
\gamma A_{1}\\
\vdots\\
\gamma^{t-1} A_{t-1}
\end{pmatrix}.~~~~~~~(3.1)$$
We associate to $G$ the matrix
$$~~~~~~~~~~~~~~~~~~~~~~~~~~~~~~~~~~~~~~~~~~A=\begin{pmatrix}
A_{0}\\
 A_{1}\\
\vdots\\
 A_{t-1}
\end{pmatrix}.~~~~~~~~~~~~~~~~~~~~~~~~~~~~~~~~~~~~~~~~~(3.2)$$
\end{defi}
For a code $C$, we define the rank of $C$, denoted by rank$(C)$, to be the minimum
number of generators of $C$ and the free rank of $C$, denoted by free rank$(C)$,
to be the maximum of the ranks of free $R$-submodules of $C$. Codes where the
rank is equal to the free rank are called free codes.

Let $C$ be a linear code over $R$. We denote by $k(C)$ the number of rows of a generating matrix $G$ in standard form for $C$, and for $i=1,2,\ldots,t-1$ we denote by $k_{i}(C)$ the number of rows of  $G$ that are divisible by $\gamma^{i}$ but not by $\gamma^{i+1}$.

Clearly, rank$C=k(C)=\sum_{i=0}^{t-1}k_{i}(C).$

It is well known (see [10]) that for codes $C$ of length $n$ over any alphabet of
size $m$
$$d_{H}(C)\leq n -log_{m}(|C|) + 1.$$
Codes meeting this bound are called MDS (Maximum Distance Separable) codes.

Further if C is linear, then
$$d_{H}(C) \leq n-rank(C) + 1.$$
Codes meeting this bound are called MDR (Maximum Distance with respect to Rank) codes.

\begin{lem}\label{3.4}
Let $C$ be a linear code over $R$ with a generator matrix $G$ in standard form and let $A$ be as in $(3.2)$.\\

$(1)$For $0\leq i\leq t-1$, $\overline{(C:\gamma^{i})}$ has generator matrix
$$
\overline{G_{i}}=\left(
\begin{array}{c}
\overline{A_{0}} \\
\vdots\\
\overline{A_{i}}
\end{array}
\right)
$$
and $\mathrm{dim}\overline{(C:\gamma^{i})}=k_{0}(C)+\cdots+k_{i}(C)$.\\

$(2)$ For $0\leq i\leq t-1$,  $\overline{(C^{\perp}:\gamma^{i})}=\overline{(C:\gamma^{t-1-i})}^{\perp}$. We have $k(C^{\perp})=n-k_{0}(C),k_{0}(C^{\perp})=n-k(C)$, and $k_{i}(C^{\perp})=k_{t-i}(C)$, for $i=1,\ldots,t-1$.

$(3)$ $d_H(C)=d_H\overline{(C^{\perp}:\gamma^{t-1})}$.

$(4)$ If $C$ is an MDR code over $R$, then $\overline{(C^{\perp}:\gamma^{t-1})}$ is an MDS code over $\mathbb{F}_q=R/\langle\gamma\rangle$.
\end{lem}

We have an important observation that proves to be rather useful to construct SQC .

\begin{lem}
Let $C$ be a self-dual code of length $n$ over finite chain ring $R$. Then $$\overline{(C:\gamma^{t-1-i})}^{\perp}\subseteq\overline{(C:\gamma^{i+j})},$$ where $0\leq i\leq t-1,~0\leq j\leq t-1-i$. In particular, $$\overline{(C:\gamma^{t-1})}^{\perp}\subseteq\overline{(C:\gamma^{t-1})}.$$
\end{lem}
\pf For $1\leq i\leq t-1,~0\leq j\leq t-1-i$, by  definition 3.2 and Lemma 3.4,  we have
$$\overline{(C:\gamma^{t-1-i})}^{\perp}=\overline{(C^{\perp}:\gamma^{i})}=\overline{(C:\gamma^{i})}\subseteq\overline{(C:\gamma^{i+j})}.$$
In case $i=0$, obviously,
$\overline{(C:\gamma^{t-1})}^{\perp}\subseteq\overline{(C:\gamma^{t-1})}$.
\qed
\begin{thm}
Let $C$ be a self-dual code of length $n$ and minimum distance $d_H(C)$ over finite chain ring $R$ with a generator matrix $G$ in standard form. Then

$(1)$ there exists  a  quantum code with parameters $[[n, 2k(C)-n,\geq d_H(C)]]_{q}$. In particular, if $C$ is an MDR code, then there exists  an  SQCMDS code with parameters $[[n, 2k(C)-n, d_H(C)]]_{q}$.

$(2)$ there exists a quantum code with parameters $[[n, l+2s-n,\geq d_1]]_{q}$,  where $ d_1=\mathrm{min}\{d_H\overline{(C:\gamma^{t-1-i})}, d_H\overline{(C:\gamma^{i+j})}\}$, $s=k_{0}(C)+k_{1}(C)+\cdots+k_{i}(C)$,$l=k_{i+1}(C)+\cdots+k_{i+j}(C)$, and $ 0\leq i\leq t-1, 0\leq j \leq t-1-i$.
\end{thm}
\pf By Lemma 3.4 (1), We know that $\mathrm{dim}\overline{(C:\gamma^{t-1})}=k(C)$. Thus,
there exists a $[n,k(C),d_H(C)] $ code with $\overline{(C:\gamma^{t-1})}^{\perp}\subseteq\overline{(C:\gamma^{t-1})}$.
 According Corollary 2.3,  the part (1) is proved.

For (2), by Lemma 3.4 (2), $\mathrm{dim}\overline{(C:\gamma^{t-1-i})}^{\perp}=k_0(C)+k_{1}(C)+\cdots+k_{i}(C)$, and $\mathrm{dim}\overline{(C:\gamma^{i+j})}=k_{0}(C)+k_{1}(C)+\cdots+k_{i}(C)+\cdots+k_{i+j}(C)$. Using Theorem 2.2 and Lemma 3.5, there exists a quantum code with parameters $[[n, l+2s-n,\geq d_1]]_{q}$,  which is the required result.
\qed

In the rest of this section, we aim to obtain good quantum codes by cyclic codes over a finite chain ring $\widetilde{R}$ with maximal ideal $\mathfrak{m}=\widetilde{R}\gamma$, where $\gamma$ is a generator of  $\mathfrak{m}$ with nilpotency index $2$.

The following  result is well known (see [11]).
\begin{thm}
Let $C$ be a cyclic code of length $n$  over finite chain ring $\widetilde{R}$ with characteristic $p^a$, where $(p,n)=1$.  Then

$(1)$ $C=\langle f(x)h(x),\gamma f(x)g(x)\rangle$,  where $f(x)g(x)h(x)=x^n-1$.

$(2)$ $C^{\perp}=\langle g^{*}(x)h^{*}(x),\gamma g^{*}(x)f^{*}(x)\rangle$, where $g^{*}(x)=x^{\mathrm{deg}g(x)}g(\frac{1}{x})$, i.e., $g^{*}(x)$ is the reciprocal of $g(x)$.

$(3)$ $\overline{C}=\langle\overline{fh}\rangle$, and $\overline{(C:\gamma)}=\langle\overline{f}\rangle$.
\end{thm}
\begin{thm}
Let $C$ be a cyclic code of length $n$  over finite chain ring $\widetilde{R}$ with characteristic $p^a$, where $(p,n)=1$. If $C=\langle f(x)h(x),\gamma f(x)g(x)\rangle$ with $f(x)g(x)h(x)=x^n-1$, then $C$ is self-dual if and only if $f(x)= \epsilon g^{*}(x)$ and $h(x)= \varepsilon h^{*}(x)$, where $\epsilon$ and $\varepsilon$ are units.
\end{thm}
\pf  The sufficiency is obvious since $C^{\perp}=\langle g^{*}(x)h^{*}(x),\gamma g^{*}(x)f^{*}(x)\rangle$.

Now, If $C$ is self-dual, by Theorem 3.7 (2)  we know that $\langle f(x)h(x),\gamma f(x)g(x)\rangle=\langle g^{*}(x)h^{*}(x),\gamma g^{*}(x)f^{*}(x)\rangle$. But these generators are the unique generators of this form. Hence
$$f(x)h(x)= g^{*}(x)h^{*}(x).$$
and
$$f(x)h(x)g(x)= g^{*}(x)h^{*}(x)g(x)=-g^{*}(x)h^{*}(x)f^{*}(x)=x^n-1.$$
Since $f^{*}(x)$ and $g^{*}(x)h^{*}(x)$ are coprime, $f^{*}(x)\mid g(x)$. Similarly, since
$$f(x)h(x)f^{*}(x)= g^{*}(x)h^{*}(x)f^{*}(x)=-f(x)g(x)h(x).$$
and $g(x)$ and $f(x)h(x)$ are coprime, $g(x)\mid f^{*}(x) $. That means that $g(x)=\epsilon f^{*}(x)$.
Now, $f(x)h(x)=g^{*}(x)h^{*}(x)=\epsilon f(x)h^{*}(x)$ where $h^{*}(x)$ and $f(x)$ are coprime. This implies that $h(x)\mid h^{*}(x)$. Similarly, since $f^{*}(x)h^{*}(x)=g(x)h(x)=\epsilon f^{*}(x)h(x)$ where $h^{*}(x)$ and  $f^{*}(x)$ are coprime,  $h^{*}(x)\mid h(x)$. Therefore, $h(x)= \varepsilon h^{*}(x)$.
\qed

Now combining Theorem 3.6 , 3.7 and 3.8, the following result is obtained.
\begin{thm}
Let $C=\langle f(x)h(x),\gamma f(x)g(x)\rangle$ be a cyclic self-dual code of length $n$  over finite chain ring  $\widetilde{R}$ with characteristic $p^a$, where $(p,n)=1$ and $f(x)g(x)h(x)=x^n-1$.  Then

$(1)$ there exists a quantum code with parameters $[[n,n-2\mathrm{deg}\overline{f(x)}, \geq d_H\overline{(C:\gamma)}~]]_q$.

$(2)$ there exists a quantum code with parameters $[[n,n-2\mathrm{deg}\overline{f(x)}-\mathrm{deg}\overline {h(x)}, \geq d_H\overline{(C:\gamma)}~]]_q$.
\end{thm}

\begin{exam} We first provide some examples  to obtain  quantum codes by non-trivial self-dual cyclic  codes over the chain ring  $\mathbb{F}_2+u\mathbb{F}_2$, where $u^2=0$.

$(1)$Let $n=7$. The factorization of $x^7+1$ over  $\mathbb{F}_2+u\mathbb{F}_2$:
$$x^7-1=(x+1)(x^3+x^2+1)(x^3+x+1).$$
Let $f(x)=x^3+x^2+1, g(x)=x^3+x+1,h(x)=x+1$. Then by Theorem 3.8  cyclic  code  $C=\langle f(x)h(x),uf(x)g(x)\rangle$ is self-dual. Using Theorem 3.9 (1), a $[[7,1,\geq3]]_2$ quantum code may be obtained from the $\overline{(C:u)}=\langle\overline{f}\rangle$ of this code. Obviously, it is a SQCNMDS code. Again using Theorem 3.9 (2), a $[[7,0,\geq3]]_2$ quantum code may be obtained from the $\overline{C}=\langle\overline{f(x)h(x)}\rangle$ of this code.

$(2)$ Let $n=15$. The factorization of $x^{15}+1$ over  $\mathbb{F}_2+u\mathbb{F}_2$  is equal to$f_1f_2f_3f_4f_5$, where $f_1=x+1,f_2=x^2+x+1,f_3=x^4+x+1,f_4=x^4+x^3+1,f_5=x^4+x^3+x^2+x+1$.
Set $f(x)=f_3, g(x)=f_4,h(x)=f_1f_2f_5$. Then by Theorem 3.8  cyclic  code  $C=\langle f(x)h(x),uf(x)g(x)\rangle$ is self-dual. Using Theorem 3.9 (1), a $[[15,7,\geq3]]_2$ quantum code may be obtained from the $\overline{(C:u)}=\langle\overline{f}\rangle$ of this code.  Again using Theorem 3.9 (2), a $[[15,0,\geq3]]_2$ quantum code may be obtained from the $\overline{C}=\langle\overline{f(x)h(x)}\rangle$ of this code.

$(3)$ Let $n=21$. The factorization of $x^{21}+1$ over $\mathbb{F}_2+u\mathbb{F}_2$ is equal to $f_1f_2f_3f_4f_5f_6$, where $f_1=x+1,f_2=x^2+x+1,f_3=x^3+x+1,f_4=x^3+x^2+1,f_5=x^6+x^4+x^2+x+1,f_6=x^6+x^5+x^4+x^2+1$.
Set $f(x)=f_5, g(x)=f_6,h(x)=f_1f_2f_3f_4$. Then by Theorem 3.8  cyclic  code  $C=\langle f(x)h(x),uf(x)g(x)\rangle$ is self-dual. Using Theorem 3.9 (1), a $[[21,9,\geq3]]_2$ quantum code may be obtained from the $\overline{(C:u)}=\langle\overline{f}\rangle$ of this code.  Again using Theorem 3.9 (2), a $[[21,0,\geq3]]_2$ quantum code may be obtained from the $\overline{C}=\langle\overline{f(x)h(x)}\rangle$ of this code.

$(4)$ Let $n=23$. The factorization of $x^{23}+1$ over  $\mathbb{F}_2+u\mathbb{F}_2$ is equal to $f_1f_2f_3$, where $f_1=x+1,f_2=x^{11}+x^{9}+x^{7}+x^{6}+x^{5}+x+1,f_3=x^{11}+x^{10}+x^{6}+x^{5}+x^{4}+x^2+1$.
Set $f(x)=f_2, g(x)=f_3,h(x)=f_1$. Then by Theorem 3.8  cyclic  code  $C=\langle f(x)h(x),uf(x)g(x)\rangle$ is self-dual. Using Theorem 3.9 (1), a $[[23,1,\geq7]]_2$ quantum code may be obtained from the $\overline{(C:u)}=\langle\overline{f}\rangle$ of this code.  Again using Theorem 3.9 (2), a $[[23,0,\geq7]]_2$ quantum code may be obtained from the $\overline{C}=\langle\overline{f(x)h(x)}\rangle$ of this code.

$(5)$ Let $n=31$. The factorization of $x^{31}+1$ over  $\mathbb{F}_2+u\mathbb{F}_2$ is equal to $(x+1)f_1f_{1c}f_2f_{2c}f_3f_{3c}$, where $f_1=x^{5}+x^{2}+1,f_{1c}=x^{5}+x^{3}+1,f_2=x^{5}+x^{3}+x^{2}+x+1,f_{2c}=x^{5}+x^{4}+x^{3}+x^{2}+1,f_3=x^{5}+x^{4}+x^{2}+x+1,f_{3c}=x^{5}+x^{4}+x^{3}+x+1$.
Set $f(x)=f_1f_2, g(x)=f_{1c}f_{2c},h(x)=(x+1)f_3f_{3c}$. Then by Theorem 3.8  cyclic  code  $C=\langle f(x)h(x),uf(x)g(x)\rangle$ is self-dual. Using Theorem 3.9 (1), a $[[31,21,\geq5]]_2$ quantum code may be obtained from the $\overline{(C:u)}=\langle\overline{f}\rangle$ of this code.  Obviously, it is a SQCNMDS code.  Again using Theorem 3.9 (2), a $[[31,0,\geq5]]_2$ quantum code may be obtained from the $\overline{C}=\langle\overline{f(x)h(x)}\rangle$ of this code.
\end{exam}

\begin{exam} We provide some examples  to obtain  quantum codes by non-trivial cyclic self-dual codes over the chain ring  $\mathbb{Z}_{p^{2}}$.

$(1)$ Length $11$ over $\mathbb{Z}_{3^{2}}$.
The factorization of $x^{11}-1$ over $\mathbb{Z}_{3^{2}}$ into a product of basic irreducible polynomials over $\mathbb{Z}_{3^{2}}$ is given by
$$x^{11}-1=f_1(x)f_2(x)f_3(x).$$
where $f_1=x-1,f_2=x^5+3x^4+8x^3+x^2+2x-1,f_3=x^5-2x^4-x^3+x^2-3x-1$. Let $f(x)=-f_3(x), g(x)=f_1(x),h(x)=-f_1(x)$. Then by Theorem 3.8  cyclic  code  $C=\langle f(x)h(x),3f(x)g(x)\rangle$ is self-dual. Using Theorem 3.9 (1), a $[[11,1,\geq2]]_3$ quantum code may be obtained from the $\overline{(C:3)}=\langle\overline{f}\rangle$ of this code.  Again using Theorem 3.9 (2), a $[[11,0,\geq2]]_3$ quantum code may be obtained from the $\overline{C}=\langle\overline{f(x)h(x)}\rangle$ of this code.

Length $13$ over $\mathbb{Z}_{3^{2}}$.
The factorization of $x^{13}-1$ over $\mathbb{Z}_{3^{2}}$ into a product of basic irreducible polynomials over $\mathbb{Z}_{3^{2}}$ is given by
$$x^{13}-1=f_1(x)f_2(x)f_3(x)f_4(x)f_5(x).$$
where $f_1=x-1,f_2=x^3+6x^2+2x+8,f_3=x^3+7x^2+3x+8,f_4=x^3+4x^2+7x+8,f_5=x^3+2x^2+7x+8$. Let $f(x)=f_3(x), g(x)=f_2(x),h(x)=-f_1(x)f_4(x)f_5(x)$. Then by Theorem 3.8  cyclic  code  $C=\langle f(x)h(x),3f(x)g(x)\rangle$ is self-dual. Using Theorem 3.9 (1), a $[[13,7,\geq3]]_3$ quantum code may be obtained from the $\overline{(C:3)}=\langle\overline{f}\rangle$ of this code.  Obviously, it is a SQCNMDS code. Again using Theorem 3.9 (2), a $[[13,0,\geq3]]_3$ quantum code may be obtained from the $\overline{C}=\langle\overline{f(x)h(x)}\rangle$ of this code.

$(2)$ Length $11$ over $\mathbb{Z}_{5^{2}}$.
The factorization of $x^{11}-1$ over $\mathbb{Z}_{5^{2}}$ into a product of basic irreducible polynomials over $\mathbb{Z}_{5^{2}}$ is given by
$$x^{11}-1=f_1(x)f_2(x)f_3(x).$$
where $f_1=x-1,f_2=x^5+17x^4+24x^3+x^2+16x+24,f_3=x^5+9x^4+24x^3+x^2+8x+24$. Let $f(x)=-f_3(x), g(x)=f_2(x),h(x)=-f_1(x)$. Then by Theorem 3.8  cyclic  code  $C=\langle f(x)h(x),5f(x)g(x)\rangle$ is self-dual. Using Theorem 3.9 (1), a $[[11,1,\geq5]]_5$ quantum code may be obtained from the $\overline{(C:5)}=\langle\overline{f}\rangle$ of this code. Obviously, it is a SQCNMDS code. Again using Theorem 3.9 (2), a $[[11,0,\geq5]]_5$ quantum code may be obtained from the $\overline{C}=\langle\overline{f(x)h(x)}\rangle$ of this code.

Length $22$ over $\mathbb{Z}_{5^{2}}$.
The factorization of $x^{22}-1$ over $\mathbb{Z}_{5^{2}}$ into a product of basic irreducible polynomials over $\mathbb{Z}_{5^{2}}$ is given by
$$x^{22}-1=f_1(x)f_2(x)f_3(x)f_4(x)f_5(x)f_6(x).$$
where $f_1=x-1,f_2=x+24,f_3=x^5+16x^4+24x^3+24x^2+8x+1,f_4=x^5+17x^4+24x^3+x^2+16x+1,f_5=x^5+8x^4+24x^3+24x^2+16x+1,f_6=x^5+9x^4+24x^3+x^2+16x+1$. Let $f(x)=f_5(x), g(x)=f_3(x),h(x)=f_1(x)f_2(x)f_4(x)f_6(x)$. Then by Theorem 3.8  cyclic  code  $C=\langle f(x)h(x),5f(x)g(x)\rangle$ is self-dual. Using Theorem 3.9 (1), a $[[22,12,\geq2]]_5$ quantum code may be obtained from the $\overline{(C:5)}=\langle\overline{f}\rangle$ of this code.  Again using Theorem 3.9 (2), a $[[22,0,\geq3]]_5$ quantum code may be obtained from the $\overline{C}=\langle\overline{f(x)h(x)}\rangle$ of this code.

$(3)$ Length $6$ over $\mathbb{Z}_{7^{2}}$.
The factorization of $x^{6}-1$ over $\mathbb{Z}_{7^{2}}$ into a product of basic irreducible polynomials over $\mathbb{Z}_{7^{2}}$ is given by
$$x^{6}-1=f_1(x)f_2(x)f_3(x)f_4(x)f_5(x)f_6(x).$$
where $f_1=x-1,f_2=x+1,f_3=x+18,f_4(x)=x-18,f_5(x)=x+19,f_6(x)=x-19$. Let $f(x)=-18f_6(x), g(x)=f_4(x),h(x)=-18f_1(x)f_2(x)f_3(x)f_5(x)$. Then by Theorem 3.8  cyclic  code  $C=\langle f(x)h(x),7f(x)g(x)\rangle$ is self-dual. Using Theorem 3.9 (1), a $[[6,4,\geq2]]_7$ quantum code may be obtained from the $\overline{(C:7)}=\langle\overline{f}\rangle$ of this code. Obviously, it is a SQCMDS code. Again using Theorem 3.9 (2), a $[[6,0,\geq2]]_7$ quantum code may be obtained from the $\overline{C}=\langle\overline{f(x)h(x)}\rangle$ of this code.
\end{exam}

\section{SQC  from to the Cyclic Codes over  Chain Rings $\mathbb{F}_{p^{2m}}+u\mathbb{F}_{p^{2m}}$ }
Throughout this section, $p$ denotes a prime number and $\mathbb{F}_{p^{2m}}$ denotes the finite field with $p^{2m}$ elements for a positive integer $m$. We always assume that $n$ is a positive integer.

The ring $R= \mathbb{F}_{p^{2m}}+u\mathbb{F}_{p^{2m}}$ consists of all $p^{2m}$-ary polynomials of degree $0$ and $1$ in an indeterminate $u$, and it is closed under $p^{2m}$-ary polynomial addition and multiplication  modulo $u^2$. Thus, $R=\frac{ \mathbb{F}_{p^{2m}}[u]}{\langle u^2\rangle}=\{a+ub|a,b\in\mathbb{F}_{p^{2m}}\}$ is a local ring with maximal ideal $u\mathbb{F}_{p^{2m}}$. Therefore, it is a chain ring. The ring $R$ has precisely $p^{2m}(p^{2m}-1)$ units, which are  of the forms $\alpha+u\beta$ and $\gamma$, where $\alpha,\beta$, and $\gamma$ are  nonzero elements of the field $\mathbb{F}_{p^{2m}}$.

Let $\overline{a+ub}:=\overline{a}+u\overline{b}$, where $\overline{a}=a^{p^m}$, and $\overline{b}=b^{p^m}$. The  Hermitian inner product over  $\mathbb{F}_{p^{2m}}+u\mathbb{F}_{p^{2m}}$ is defined as follows:
$$[\mathbf{x},\mathbf{y}]_H=\sum_{i=1}^{n}x_i\overline{y_i}.$$
where $\mathbf{x},\mathbf{y}\in R^n$, $\mathbf{x}=(x_1,\ldots,x_n)$ and $\mathbf{y}=(y_1,\ldots,y_n)$. The Hermitian dual code $C^{\perp_H} $ of $C$ is defined by
$$C^{\perp_H}=\{\mathbf{x}\in R^n~|~[\mathbf{x},\mathbf{y}]_H=0~\mathrm{for~all}~~\mathbf{y}~\in C\}.$$

It is evident that $C^{\perp_H}$ is linear. We say that a code $C$ is Hermitian dual-containing code if $C^{\perp_H}\subset C$ and $C\neq R^n$, and Hermitian self-dual if $C^{\perp_H}=C$.

It is easy to prove that
$$~~~~~~~~~~~~~~~~~~~\mid C\mid \cdot \mid C^{\perp_H}\mid=\mid R\mid^{n}.~~~~~~~~~~~~~~~~~~~~~~~~~~~~(4.1)$$

The following lemma can be found in  \cite{Dougherty2010}.
\begin{lem}  $[ 12,~  Corollary~4.2]$  Assume the notations given above. Then there exist $\alpha\in R$ such that $\alpha^2=-1$ if and only if $p^{2m}\equiv1~(\mathrm{mod}~4)$.

\end{lem}
\begin{rem} Since  $\alpha\in R$, there exist $s,t\in \mathbb{F}_{p^{2m}}$ such that $\alpha=s+ut$. Hence computing in $R$, we have $\alpha^2=s^2+2stu=-1$, which implies that $s^2=-1,2st=0$. If $p=2$, then take $s=1\in \mathbb{F}_{p^{2m}},t=0$ we have $\alpha^2=-1$; If $p\neq2$, then $t=0$ since $2st=0$. Therefore, $\alpha=s\in \mathbb{F}_{p^{2m}}$.
\end{rem}

From now on, we always assume that $p^{m}\equiv1~(\mathrm{mod}~4)$, then  $p^{2m}\equiv1~(\mathrm{mod}~4)$.  So there exist $\alpha \in \mathbb{F}_{p^{2m}}$ such that $\alpha^2=-1$ in $R$.

We first give the definition of the Gray map on $R^n$. The Gray map $\Phi_1:~R\rightarrow\mathbb{F}_{p^{2m}}^{2}$ is given by $\Phi_1(a+bu)=(\alpha b,a+b)$, where $\alpha^2=-1$. This map can be extended to $R^n$ in a natural way:
\begin{eqnarray*}
\Phi:&R^n\longrightarrow \mathbb{F}_{p^{2m}}^{2n}\\
     &(a_1+ub_1,\ldots,a_{n}+ub_{n})\longmapsto (\alpha b_{1},a_1+b_1,\ldots,\alpha b_n,a_n+b_{n}).
\end{eqnarray*}

Next, we define a Gray weight for codes over $R$ as follows.
\begin{defi} The Gray weight over $R$ is a weight function on $R$ defined as:
$$w_{G}(a+bu)=\left\{\begin{aligned}
         &0~~ ~~~~~~~if~a=0,~b=0,\\
         &1~~~~~ ~~~~~if~a\neq0,~b=0,\\
         &1~~~~~ ~~~~~if~b\neq0,a+b\equiv0~(\mathrm{mod~}p),\\
         &2~~~~~ ~~~~~if~b\neq0,a+b\neq0~(\mathrm{mod~}p).
         \end{aligned}\right. $$
Define the Gray weight of a codeword $\mathbf{c}=(c_1,\ldots,c_n)\in R^n$ to be the rational sum of the Gray weight of its components is, $w_G(\mathbf{c})=\sum_{i=1}^{n}w_G(c_i)$. For any $\mathbf{c}_1,\mathbf{c}_2\in R^n$, the Gray distance $d_G$ is given by $d_G(\mathbf{c}_1,\mathbf{c}_2)=w_G(\mathbf{c}_1-\mathbf{c}_2)$. The minimum Gray distance of $C$ is the smallest nonzero Gray distance between all pairs of distinct codewords of $C$. The minimum Gray weight of $C$ is the smallest nonzero Gray weight among all  codewords of $C$. If $C$ is linear, then the minimum Gray distance is same as the minimum Gray weight.

The Hamming weight $W(\mathbf{c})$ of a codeword $\mathbf{c}$ is the number of nonzero components in $\mathbf{c}$. The Hamming distance $d(\mathbf{c}_1,\mathbf{c}_2)$ between two codewords $\mathbf{c}_1$ and $\mathbf{c}_2 $ is Hamming weight of the codeword $\mathbf{c}_1-\mathbf{c}_2$. The minimum Hamming distance $d$ of $C$ is defined as $\mathrm{min}\{d(\mathbf{c}_1,\mathbf{c}_2)|\mathbf{c}_1,\mathbf{c}_2\in C,\mathbf{c}_1\neq\mathbf{c}_2\}$ (See $\mathrm{[13]}$).
\end{defi}
\begin{prop} The Gray map $\Phi$ is a distance-preserving map from $(R^n,~\mathrm{Gray~distance})$ to $(\mathbb{F}_{p^{2m}}^{2n},~\mathrm{Hamming~distance})$ and it is also $\mathbb{F}_{p^{2m}}$-linear.
\end{prop}
\pf From the above definitions, it is clear that $\Phi(\mathbf{c}_1-\mathbf{c}_2)=\Phi(\mathbf{c}_1)-\Phi(\mathbf{c}_2)$ for $\mathbf{c}_1,\mathbf{c}_2\in R^n$. Thus,
$$d_G(\mathbf{c}_1,\mathbf{c}_2)=w_G(\mathbf{c}_1-\mathbf{c}_2)=w(\Phi(\mathbf{c}_1-\mathbf{c}_2))=w(\Phi(\mathbf{c}_1)-\Phi(\mathbf{c}_2))=d(\Phi(\mathbf{c}_1),\Phi(\mathbf{c}_2)).$$

Let $\mathbf{c}_1,\mathbf{c}_2\in R^n$, $k_1,k_2\in\mathbb{F}_{p^{2m}}$. From the definition of the Gray map, we have $\Phi(k_1\mathbf{c}_1+k_2\mathbf{c}_2)=k_1\Phi(\mathbf{c}_1)+k_2\Phi(\mathbf{c}_2)$, which means that $\Phi$ is an $\mathbb{F}_{p^{2m}}$-linear map.
\qed
 \begin{cor} If $C$ is a linear code over $R$ of length $n$,  size $(p^{2m})^k $ and minimum Gray weight $d_G$, then $\Phi(C)$ is a linear code over $\mathbb{F}_{p^{2m}}$ with parameters $[2n,k,d_G]$.
 \end{cor}
The Hermitian inner product over $\mathbb{F}_{p^{2m}}$ is defined as follows:
 $$[\mathbf{a},\mathbf{b}]_H=\mathbf{a}\cdot \overline{\mathbf{b}}=\sum_{i=1}^{n}a_{i}b_{i}^{p^m},$$
where  $\mathbf{a},\mathbf{b}\in\mathbb{F}_{p^{2m}}^{n},\mathbf{a}=(a_1,\ldots,a_n),\mathbf{b}=(b_1,\ldots,b_n)$ and $\cdot$ is the usual Euclidean inner product.

An important connection that we want to investigate is the relation between the Hermitian dual and the Gray image of a code. We have the following theorem.
\begin{thm}Let $C$ be a linear code over $R$ of length $n$. Then $\Phi(C^{\perp_H})=\Phi(C)^{\perp_H}$.
\end{thm}
\pf To  prove the theorem, we first show $\Phi(C^{\perp_H})\subset\Phi(C)^{\perp_H}$,  i.e., $$[\mathbf{x},\mathbf{y}]_{H}=0~\Rightarrow~[\Phi(\mathbf{x}),\Phi(\mathbf{y})]_H=0~~\mathrm{for~all}~\mathbf{x},\mathbf{y}\in R^n.~~~~~~~~~~~~~~~~~~~(4.2)$$

To this extent, let us assume that $\mathbf{x}=(a_{1}+ub_1,\ldots,a_n+ub_n)$ and $\mathbf{y}=(c_{1}+ud_1,\ldots,c_n+ud_n)$,  where $a_i,b_i,c_i,d_i\in\mathbb{F}_{p^{2m}}$. Then by
$$[\mathbf{x},\mathbf{y}]_{H}=\sum_{i=1}^{n}a_i\overline{c_i}+\sum_{i=1}^{n}(b_i\overline{c_i}+a_i\overline{d_i})u,$$
we see that $[\mathbf{x},\mathbf{y}]_{H}=0$ if and only if
$$~~~~~~~~~~~~~~~~~~~~~~~~\sum_{i=1}^{n}a_i\overline{c_i}=0,~~~~~~~~~~~~~~~~~~~~~~~~~~~~~~~~~~~~~~~~~~~~~~~(4.3)$$
and
$$~~~~~~~~~~~~~~~~~~~~~~\sum_{i=1}^{n}(b_i\overline{c_i}+a_i\overline{d_i})=0.~~~~~~~~~~~~~~~~~~~~~~~~~~~~~~~~~~~~~~(4.4)$$

Note that $ p^m\equiv1~(\mathrm{mod}~4)$ we can assume that  $p^m=4k+1$ for some $k \in\mathbb{N}$,  hence, $ p^m+1=4k+2=2(2k+1)$.  According to $\alpha^2=-1$, we have
$$~~~~~~~~~~~~~~~~~~~~~~~~~~~~\alpha^{p^m+1}=(\alpha^2)^{2k+1}=-1.~~~~~~~~~~~~~~~~~~~~~~~~~~~~~~~~~~(4.5)$$

Now, since $\Phi(\mathbf{x})=(\alpha b_1,a_1+b_1,\ldots,\alpha b_n,a_n+b_n)$, and $\Phi(\mathbf{y})=(\alpha d_1,c_1+d_1,\ldots,\alpha d_n,c_n+d_n)$, we get
$$[\Phi(\mathbf{x}),\Phi(\mathbf{y})]_H=\sum_{i=1}^{n}\alpha^{p^m+1}b_i\overline{d_i}+\sum_{i=1}^{n}(a_i+b_i)(\overline{c_i+d_i})~~~~~~$$
$$~~~~~=\sum_{i=1}^{n}\alpha^{p^m+1}b_i\overline{d_i}+\sum_{i=1}^{n}(a_i\overline{c_i}+a_i\overline{d_i}+b_i\overline{c_i}+b_i\overline{d_i})$$
$$~~~~~~~~=\sum_{i=1}^{n}(\alpha^{p^m+1}+1)b_i\overline{d_i}+\sum_{i=1}^{n}a_i\overline{c_i}+\sum_{i=1}^{n}(b_i\overline{c_i}+a_i\overline{d_i}),$$
by(4.3-4.5) which finishes the of (4.2), i.e.,
   $$\Phi(C^{\perp_H})\subset\Phi(C)^{\perp_H}.~~~~~~~~~~~~~~~~~~~~~~~~~~~~~~~(4.6)$$

In the light of Corollary 4.5, $\Phi(C)$ is a linear code of length $2n$ of size $|C|$ over $\mathbb{F}_{p^{2m}}$. So, by Corollary 4.5, we know that $$|\Phi(C)^{\perp_{H}}|=\frac{(p^{2m})^{2n}}{|\Phi(C)|}=\frac{(p^{2m})^{2n}}{|C|}.$$

Since $R$ is a finite chain ring, i.e., Frobenius ring, we have
$$|C^{\perp_{H}}|\cdot|C|=|R|^{n}=(p^{2m})^{2n}.$$
Hence, this implies that
$$|\Phi(C^{\perp_{H}})|=|\Phi(C)^{\perp_{H}}|.~~~~~~~~~~~~~~~~~~~~~~~~~~~~~~~(4.6)$$
Combining (4.5) with (4.6), we get the desired equality.
\qed

The following is an immediate corollary to this:
\begin{cor}(1) If $C$ is a Hermitian self-dual code of length $n$ over $R$, then $\Phi(C)$ is a Hermitian self-dual code of length $2n$ over $\mathbb{F}_{p^{2m}}$;

(2) If $C$ is a Hermitian dual-containing code of length $n$ over $R$, then $\Phi(C)$ is a Hermitian dual-containing code of length $2n$ over $\mathbb{F}_{p^{2m}}$.
\end{cor}

In the folowing, we always assume that $n$ is a positive integer and $(n,p)=1$. Let $\mathcal{R}_{n}=\frac{R[x]}{<x^n-1>}$. We denote by $ \mu  $ the natural surjective ring morphism from $R$ to $\mathbb{F}_{p^{2m}}$, which can be extended naturally to a surjective ring morphism from $R[x]$ to $\mathbb{F}_{p^{2m}}[x]$.

For a polynomial $f(x)$ of degree $k$ in $R[x]$, its reciprocal polynomial $x^kf(x^{-1})$ is denoted by $f^{*}(x)$. Note that  the roots of $f^{*}(x)$ are the reciprocal of the corresponding roots of $f(x)$. Set $f(x)=a_{0}+a_{1}x+\cdots+a_{k}x^{k}$, we define
$$\overline{f(x)}=\overline{a}_{0}+\overline{a}_{1}x+\cdots+\overline{a}_{k}x^{k}.$$

\begin{lem} Let $f(x)=(t_{0}+us_{0})+(t_{1}+us_{1})x+\cdots+(t_{n-1}+us_{n-1})x^{n-1}\in R[x]$ and $\eta$ be a primitive $n$th root of unity in some extension ring of $R$. If $\eta^{s}$ is a root of $f(x)$, there $\overline{f^{*}(x)}$ has  $\eta^{-p^{m}s}$ as a root.
\end{lem}
\pf Let $\xi=\eta^{s}$. Then
 $$f(\xi)=(t_{0}+t_{1}\xi+\cdots+t_{n-1}\xi^{n-1})+u(s_{0}+s_{1}\xi+\cdots+s_{n-1}\xi^{n-1})=0.$$
This implies
$$t_{0}+t_{1}\xi+\cdots+t_{n-1}\xi^{n-1}=0~~\mathrm{and}~~s_{0}+s_{1}\xi+\cdots+s_{n-1}\xi^{n-1}=0.$$
Therefore,
$$\overline{f^{*}(\xi^{-p^{m}})}=(\overline{t}_{n-1}+\overline{t}_{n-2}\xi^{-p^{m}}+\cdots+\overline{t}_{0}\xi^{-p^{m}(n-1)})+u(\overline{s}_{n-1}+\overline{s}_{n-2}\xi^{-p^{m}}+\cdots+\overline{s}_{0}\xi^{-p^{m}(n-1)})$$
$$=\xi^{-p^{m}(n-1)}[(t_{0}+t_{1}\xi+\cdots+t_{n-1}\xi^{n-1})^{p^{m}}+u(s_{0}+s_{1}\xi+\cdots+s_{n-1}\xi^{n-1})^{p^{m}}]=0$$
\qed

Let $i$ be an integer such that $0\leq i\leq n-1$, and let $l$ be the smallest positive integer such that $i(p^{2m})^{l}\equiv i~\mathrm{(mod}~n)$. Then $C_i=\{i,ip^{2m},\ldots,i(p^{2m})^{l-1}\}$ is the $p^{2m}-$cyclotomic coset module $n$ containing $i$. A cyclotomic coset $C_i$ is called symmetric if $n-p^mi\in C_i$ and asymmetric otherwise. Let $I_1$ and $I_2$ be  sets of symmetric and asymmetric coset representatives modulo $n$, respectively. Since $p$ is coprime with $n$, the irreducible factors of $x^n-1$ in $\mathbb{F}_{p^{2m}}[x]$ can be described by the $p^{2m}-$cyclotomic cosets. Suppose that $\zeta$ be a primitive $n$th root of unity over some extension field of $\mathbb{F}_{p^{2m}}$. Then $\zeta$ is also a primitive $n$th root of unity over some extension ring of $R$. Let $ m_{j}(x)$ be the minimal polynomial of $\zeta^j$ with respect to $\mathbb{F}_{p^{2m}}$. Then $ m_{j}(x)=\Pi_{i\in C_{j}}(x-\zeta^{i})$,  and $ \overline{m_{j}^{*}(x)}=\Pi_{i\in C_{-p^{m}j}}(x-\zeta^{i})$ by Lemma 4.8.  Therefore,  polynomial $x^n-1$ factors uniquely into monic irreducible polynomial in  $\mathbb{F}_{p^{2m}}[x]$ as  $x^n-1=\Pi_{j\in I_1}m_{j}(x)\Pi_{j\in I_2}m_{j}(x)m_{-p^{m}j}(x)$. By $\mathrm{Hensel}^{,}$s lemma (See [9,Theorem 4.1.1]), $x^n-(1+u)$ has a unique decomposition as a product $\Pi_{j\in I_1}M_{j}(x)\Pi_{j\in I_2}M_{j}(x)M_{-p^{m}j}(x)$ of pairwise coprime monic basic irreducible polynomials in $R[x]$ with $\mu(M_{j}(x))=m_{j}(x)$ for each $j\in I_{1} \cup I_{2}$.

The following two lemmas can be found in [14].
\begin{lem}$[14, ~Theorem~3.4]$ Let $x^n-(1+u)=\Pi_{j\in I_1\cup I_2}M_{j}(x)$ be the unique factorization of $x^n-(1+u)$ into a product of monic basic irreducible pairwise coprime polynomials  in $R[x]$. If $C$ is a cyclic code of length $n$ over $R$, then $C=\langle\Pi_{j\in I_1\cup I_2}M_{j}^{k_{j}}(x)\rangle$, where $0\leq k_j\leq 2$. In this case, $\mid C \mid=(p^{2m})^{\sum_{j\in I_1\cup I_2}(2-k_{j})~deg~M_{j}}$.
\end{lem}

\begin{lem}$[14, ~Lemma~4.2]$ Let $C=\langle\Pi_{j\in I_1\cup I_2}M_{j}^{k_{j}}(x)\rangle$ be a cyclic code of length $n$ over $R$, where the polynomials $M_j$ are the pairwise coprime monic basic irreducible factors of  $x^n-(1+u)$ in $R[x]$ and $0\leq k_j\leq 2$ for each $j\in I$. Then $C^{\perp_H}=\langle\Pi_{j\in I_1\cup I_2}\overline{M^{*}_{j}(x)}^{2-k_{j}}(x)\rangle$  and  $\mid C^{\perp_H} \mid=(p^{2m})^{\sum_{j\in I_1\cup I_2}k_{j}~deg~M_{j}}$.
\end{lem}
\begin{thm} Let $C$ be a cyclic code of length $n$ over $R$.  If  $$C=<\Pi_{j\in I_{1}}M_{j}^{k_{j}}(x)\Pi_{j\in I_{2}}M_{j}^{i_{j}}(x)M_{-p^{m}j}^{l_{j}}(x)>,$$ then $C^{\perp_H}\subset C$ if and only if  $k_j=0$ or $k_j=1$ for $j\in I_1$ and $i_j+l_j\leq2$ for $j\in I_2$.
\end{thm}
\pf According to Lemma 4.10,  we have
$$C^{\perp_H}=\langle\Pi_{j\in I_{1}}M_{j}^{2-k_{j}}(x)\Pi_{j\in I_{2}}M_{j}^{2-l_{j}}(x)M_{-p^{m}j}^{2-i_{j}}(x)\rangle.$$
Comparing with $C=\langle\Pi_{j\in I_{1}}M_{j}^{k_{j}}(x)\Pi_{j\in I_{2}}M_{j}^{i_{j}}(x)M_{-p^{m}j}^{l_{j}}(x)\rangle$, it follows that $C^{\perp_H}\subset C$ if and only if $k_j=0$ or $k_j=1$ for $j\in I_1$ and $i_j+l_j\leq2$ for $j\in I_2$.
\qed

From now on, we always assume that $n=sp^t-1$. Obviously, $(n,p)=1$. In this case, we give a method to decompose $x^n-(1+u)$ into monic basic irreducible polynomials in $R(x)$. Let $g_{1}(x),g_{2}(x),\ldots,g_{r}(x)$ be monic basic irreducible polynomials in $R(x)$ such that $x^n-1=g_{1}(x)g_{2}(x)\cdots g_{r}(x)$. Note that $(1+u)^p=1$ and $(1+u)^{sp^t}=1$, we have
$$(1+u)^{n}(x^{n}-(1+u))=(1+u)^{\sum_{i=1}^{r}deg~g_{i}}[(1+u)^{p-deg~g_{1}}g_{1}((1+u)x)]\cdots[(1+u)^{p-deg~g_{r}}g_{r}((1+u)x)]$$
$$~~~~~~~~~~~~~~~~~~~~~~=(1+u)^n[(1+u)^{p-deg~g_{1}}g_{1}((1+u)x)]\cdots[(1+u)^{p-deg~g_{r}}g_{r}((1+u)x)].$$
Therefore,
$$x^{n}-(1+u)=[(1+u)^{p-deg~g_{1}}g_{1}((1+u)x)]\cdots[(1+u)^{p-deg~g_{r}}g_{r}((1+u)x)].$$
Let $f_{i}(x)=(1+u)^{p-deg~g_{i}}g_{i}((1+u)x)$ for $1\leq i\leq r$. Then the polynomial $x^n-(1+u)$ factor uniquely into monic basic irreducible polynomials in $R(x)$ as $f_{1}(x)f_2(x)\cdots f_r(x)$.

For a code $C$ of length $n$ over $R$, their torsion and residue codes are codes over $\mathbb{F}_{p^{2m}}$, defined as follows.
$$\mathrm{Tor}(C) =\{a\in\mathbb{F}_{p^{2m}}^{n}\mid ua\in C\},~~~~\mathrm{ Res}(C) =\{a\in\mathbb{F}_{p^{2m}}^{n}\mid \exists b\in\mathbb{F}_{p^{2m}}^{n}:a+ub\in C\}.$$
The reduction modulo $u$ from $C$ to $\mathrm{Res}(C)$ is given by $\varphi: C \rightarrow \mathrm{Res}(C),\varphi(a + ub) = a$. Clearly, $\varphi$ is
well defined and onto, with $\mathrm{Ker}(\varphi) = u\mathrm{Tor}(C)$, and $\varphi(C) = \mathrm{Res}(C)$. Therefore, $|\mathrm{Res}(C)| = \frac{|C|}{| \mathrm{Tor}(C)|}$. Thus,
we obtain $|C|=|\mathrm{Res}(C)|| \mathrm{Tor}(C)|$ and the  following two theorems.
\begin{thm}   Let $C=\langle\Pi_{j\in I} M_{j}^{k_{j}}(x)\rangle$ be a cyclic code of length $n$ over $R$ where $x^n-(1+u)=\Pi_{j\in I_1}M_{j}(x)\Pi_{j\in I_2}M_{j}(x)M_{-p^{m}j}(x)$, $0\leq k_j\leq 2$, and $I=I_1\cup I_2$.  Then

$(1)$ $ \mathrm{ Res}(C) =\langle \Pi_{j\in I} [\varphi (M_{j}(x))]^{\delta_{j}}\rangle $, where $\delta_{j}=k_j$ if $k_j=1$ or $0$, and $\delta_{j}=1$ if $k_j=2$;

$(2)$ $ \mathrm{ Tor}(C) =\langle \Pi_{j\in I} [(\varphi M_{j}(x))]^{\eta_{j}}\rangle$, where $\eta_{j}=0$ if $k_j=1$ or $0$, and $\eta_{j}=1$ if $k_j=2$.

\end{thm}
\pf According to the definition of $ \mathrm{ Res}(C)$, we have  $ \mathrm{ Res}(C) =\langle \Pi_{j\in I} [(\varphi M_{j}(x))]^{k_{j}}\rangle $. Note that if $f(x)$ is a monic irreducible divisor of $x^n-1$ in $\mathbb{F}_{p^{2m}}$ and $g(x)=\frac{x^n-1}{f(x)}$, then $(f(x),g(x))=1$. So there exist $a(x),b(x) \in \mathbb{F}_{p^{2m}}[x]$ such that $a(x)f(x)+b(x)g(x)=1$ in $\mathbb{F}_{p^{2m}}[x]$. Computing in $\frac{ \mathbb{F}_{p^{2m}}[x]}{\langle x^n-1\rangle}$, we get
$$a(x)f^{2}(x)=(1-b(x)g(x))f(x)=f(x)-b(x)f(x)g(x)=f(x)-b(x)(x^n-1)=f(x).$$
Consequently, $\langle f^2(x)\rangle=\langle f(x)\rangle$. This proves the (1).

For (2), since $u \Pi_{j\in I} [(\varphi M_{j}(x))]^{\eta_{j}}=u \Pi_{j\in I} [ M_{j}(x)]^{\eta_{j}}=-\Pi_{j\in I} [ M_{j}(x)]^{\eta_{j}+1}\in C$, we have $\langle \Pi_{j\in I} [(\varphi M_{j}(x))]^{\eta_{j}}\rangle \subset \mathrm{ Tor}(C)$. By Lemma 4.9.3 and $|C|=|\mathrm{Res}(C)|| \mathrm{Tor}(C)|$, we imply
$$| \langle \Pi_{j\in I} [(\varphi M_{j}(x))]^{\eta_{j}}\rangle\mid=\mid\mathrm{Tor}(C)|.$$
Thus, $ \mathrm{ Tor}(C) =\langle \Pi_{j\in I} [(\varphi M_{j}(x))]^{\eta_{j}}\rangle$.
\qed
\begin{thm}   Let $C$ be a cyclic code of length $n$ over $R$, and let $d_1$ and $d_2$ be the minimum Hamming distances of the $\mathrm{ Res}(C)$ and $ \mathrm{ Tor}(C)$, respectively. Then  $d_G(C)=\mathrm{min}\{d_1,2d_2\}$.
\end{thm}
\pf For any nonzero codeword $c=a(x)+ub(x)\in C$, if $a(x)\neq 0$, then $a(x)\in \mathrm{ Res}(C)$. Thus $w_G(c)\geq d_1$.  Otherwise, $c=ub(x)\in \mathrm{ Tor}(C)$, hence, $w_G(c)\geq 2d_2$.  So  $d_G(C)\geq \mathrm{min}\{d_1,2d_2\}$.  On the other hand, since $u\mathrm{ Tor}(C)$ is contained in $C$, we can obtain $d _G(C)\leq2d_2$. Obviously, $\mathrm{ Res}(C)\subset C$, hence $d_1\geq d _G(C)$. It follows that $\mathrm{min}\{d_1,2d_2\}\geq d_G(C)$. This proves the expected result.
\qed

Combing Corollary 2.3, 4.5, 4.7 and Theorem 4.13, we have the following result.

\begin{thm} Let $C$ be a Hermitian dual-containing cyclic code over $R$ of length $n$ size $(p^{2m})^k $, and let $d_1$ and $d_2$ be the minimum Hamming distances of the $\mathrm{ Res}(C)$ and $ \mathrm{ Tor}(C)$.Then there exists a quantum code with parameters $[[2n, 2k-2n,\geq\mathrm{min}\{d_1,2d_2\}]]_{p^m}$.
\end{thm}
\begin{exam}
Consider cyclic codes of length $25$ over $F_{13^2}+uF_{13^2}$. In $F_{13^2}+uF_{13^2}$,
$$x^{25}-1=M_0(X)M_1(x)M_2(x)M_5(x)M_{10}(x),$$
 where
$$M_0(x)=x-(1-u),$$
$$M_1(X)=x^{10}+w^{30}(1+8u)x^5+(1+3u),$$
$$M_2(X)=x^{10}+w^{54}(1+8u)x^5+(1+3u),$$
$$M_5(X)=x^2+w^{30}(1-u)x+(1-2u),$$
$$M_{10}(X)=x^2+w^{54}(1-u)x+(1-2u).$$
Let $C=\langle M_0(x)M_1(X)^2M_{10}(X)^2\rangle$. By Theorem $4.11$, $C^{\perp_{H}}\subset C$. Using Theorem $4.13$, we find that the Gray
distance of $C$ is equal to 4. Again from Theorem $4.14$, a $[[50,42,\geq4]]_{13}$ quantum code may be obtained from Gray image of this code. This code is a SQCNMDS code.
\end{exam}
\begin{exam}
Consider cyclic codes of length $8$ over $F_{3^4}+uF_{3^4}$. In $F_{3^4}+uF_{3^4}$,
$$x^{8}-1=M_0(X)M_1(x)M_2(x)M_3(x)M_{4}(x)M_{5}(X)M_{6}(X)M_{7}(X),$$
 where
$$M_0(x)=x-(1-u),~~~$$
$$M_1(x)=x+(1-u)w^{10},$$
$$M_2(X)=x+(1-u)w^{20},$$
$$M_3(X)=x+w^{30}(1-u),$$
$$M_4(X)=x-(1-u),~~~~$$
$$M_{5}(X)=x++w^{50}(1-u),$$
$$M_{6}(X)=x++w^{60}(1-u),$$
$$M_{7}(X)=x++w^{70}(1-u).$$

Let $C=\langle M_0(x)M_1(X)^2\rangle$. By Theorem $4.11$ , $C^{\perp_{H}}\subset C$. Using Theorem $4.13$, we find that the Gray
distance of $C$ is equal to 3. Again from Theorem $4.14$, a $[[16,10,\geq3]]_{9}$ quantum code may be obtained from Gray image of this code. This code is a SQCNMDS code.
\end{exam}

\section{Conclusion}
We give two   methods to construct  quantum  codes from cyclic codes over finite chain
rings. Furthermore, the results show that cyclic codes over finite chain
rings are also a good resource of constructing quantum codes. We believe that more good quantum codes can be obtained from cyclic codes over finite chain
rings. In a future work, we will use the computer algebra system MAGMA to find more new good quantum codes.

\textbf{Acknowledgements} This work was supported by Research Funds of Hubei Province, Grant
No. D20144401.

\end{document}